\documentclass{ifacconf}

\usepackage{amsmath} 
\usepackage{amssymb}
\usepackage{color}
\usepackage[labelformat = empty,position=top]{subcaption}
\usepackage[export]{adjustbox}

\newtheorem{remark}{Remark}

\usepackage{graphicx}      
\usepackage{natbib}        
\begin{document}
\begin{frontmatter}

\title{Observer design for piecewise smooth and switched systems via contraction theory\thanksref{footnoteinfo}} 

\thanks[footnoteinfo]{Corresponding author Mario di Bernardo. E-mail address: mario.dibernardo@unina.it; tel. +39-081-7683909; fax +39-081-7683186.}

\author[Napoli]{Davide Fiore} 
\author[Napoli]{Marco Coraggio} 
\author[Napoli,Bristol]{Mario di Bernardo}

\address[Napoli]{Department of Electrical Engineering and Information Technology, University of Naples Federico II, Via Claudio 21, 80125 Naples, Italy}  
\address[Bristol]{Department of Engineering Mathematics, University of Bristol, BS8~1TR Bristol, U.K.}

\begin{abstract}                
The aim of this paper is to present the application of an approach to study contraction theory recently developed for piecewise smooth and switched systems.
The approach that can be used to analyze incremental stability properties of so-called Filippov systems (or variable structure systems) is based on the use of regularization, a procedure to make the  vector field of interest differentiable before analyzing its properties.
We show that by using this extension of contraction theory to nondifferentiable vector fields, it is possible to design observers for a large class of piecewise smooth systems using not only Euclidean norms, as also done in previous literature, but also non-Euclidean norms.
This allows greater flexibility in the design and encompasses the case of both piecewise-linear and piecewise-smooth (nonlinear) systems.
The theoretical methodology is illustrated via a set of representative examples.
\end{abstract}



\end{frontmatter}

\section{Introduction}
The problem of designing state observers for nondifferentiable systems is the subject of current research.
For example, the design of observers for Lipschitz continuous nonlinear systems was investigated in \citep{rajamani1998observers,zemouche2013lmi}, while in \citep{arcak2001observer,brogliato2009observer} design approaches based on passivity theory were proposed for Lur'e-type systems.
Also, in \citep{heemels2007observer,doris2008observer} sufficient conditions were presented to ensure stability of the estimation error for state observers of bimodal piecewise linear (PWL) systems (both continuous and discontinuous on the switching surface).
The analysis was conducted analyzing the quadri-modal estimation error dynamics based on quadratic Lyapunov functions and LMIs.
Related results were presented in \citep{van2008tracking} for the case of piecewise affine (PWA) systems.
Therein, using theoretical results developed in \citep{pavlov2007convergence}, sufficient conditions guaranteeing exponential stability of the estimation error were  given in terms of a set of appropriate LMIs.
More recently, the state estimation problem was investigated in \citep{heemels2011observer} for linear complementarity systems and in \citep{forni2013follow} for hybrid systems with impacts. \\
Contraction theory \citep{lohmiller1998contraction,russo2010global, jouffroy2005some, forni2014differential, aminzare2014contraction} is a powerful analysis tool providing sufficient conditions for incremental stability \citep{angeli2002lyapunov} of a dynamical system.
Namely, if the system vector field is contracting in a set of interest, any two of its trajectories will converge towards each other in that set, a property that can be effectively exploited to design state observers and solve tracking control problems as discussed, for instance, in \citep{lohmiller1998contraction,  van2008tracking, bonnabel2011contraction, dinh2013contraction, manchester2014control, manchester2014output, di2015switching}.
More specifically, incremental exponential stability over a given forward invariant set is guaranteed if some matrix measure, say $\mu$, of the system Jacobian matrix is uniformly negative in that set for all time. \\
The original results on contraction analysis were presented for continuously differentiable vector fields limiting their application to observer design for this class of dynamical systems.
Recently, extensions have been presented in the literature for applying contraction and convergence analysis to different classes of nondifferentiable and discontinuous vector fields \citep{lohmiller2000nonlinear, pavlov2007convergence, di2014contraction, lu2015contraction, di2013incremental, di2014incremental, fiore2015contraction}. \\
In this paper we propose a methodology to design state observers  for nondifferentiable bimodal vector fields, which stems from the results presented in \citep{fiore2015contraction} on extending contraction analysis to Filippov systems.
Specifically, we derive conditions on the observer dynamics for the estimation error to converge exponentially to zero.
These conditions, when particularized to the case of PWA systems, generalize those presented in \citep{van2008tracking} to the case of non-Euclidean norms. \\
In what follows, after reviewing some key results on contraction analysis of switched systems, we present our procedure for state observer design complementing the theoretical derivations with some illustrative examples.

%
%


\section{Contraction analysis of switched systems}
\label{sec:background}
\subsection{Incremental Stability and Contraction Theory}
Let $U\subseteq\mathbb{R}^n$ be an open set. Consider the system of ordinary differential equations
\begin{equation}
\label{eq:dynamical_sys}
\dot{x} = f(t, x),
\end{equation}
where $f$ is a continuously differentiable vector field defined for $t\in[0,\infty)$ and $x\in U$, that is $f\in C^1(\mathbb{R}^+\times U,\mathbb{R}^n)$. \\
We denote by $\psi(t,t_0,x_0)$ the value of the solution $x(t)$ at time $t$ of the differential equation \eqref{eq:dynamical_sys} with initial value $x(t_0)=x_0$.
We say that a set $\mathcal{C}\subseteq \mathbb{R}^n$ is \emph{forward invariant} for system \eqref{eq:dynamical_sys} if $x_0 \in \mathcal{C}$ implies $\psi(t,t_0,x_0) \in \mathcal{C}$ for all $t\ge t_0$.\\
A nonlinear dynamical system \eqref{eq:dynamical_sys} is \emph{contracting} if it forgets initial conditions or temporary state perturbations exponentially fast, implying convergence of system trajectories towards each other and consequently towards a steady-state solution which is determined only by the input (\emph{entrainment} property). The following theorem summarize the basic results of contraction theory \citep{russo2010global,lohmiller1998contraction}.
\begin{thm}
Let $\mathcal{C}\subseteq U$ be a forward invariant $K$-reachable set. The continuously differentiable vector field \eqref{eq:dynamical_sys} is said to be \emph{contracting} on $\mathcal{C}$ if there exists some norm $|\cdot|$ in $\mathcal{C}$, with associated matrix measure $\mu$ (see Appendix \ref{sec:appendix}), such that, for some constant $c>0$ (the \emph{contraction rate}),
\begin{equation}
\label{eq:contraction_cond}
\mu\left(\frac{\partial f}{\partial x}(t,x)\right)\leq -c, \quad \forall x\in\mathcal{C}, \; \forall t\geq t_0.
\end{equation}
Then, for every two solutions $x(t)=\psi(t,t_0,x_0)$ and $y(t)=\psi(t,t_0,y_0)$ with initial conditions $x_0,y_0\in\mathcal{C}$ we have that
\begin{equation}
\label{eq:ies}
\lvert x(t)-y(t)\rvert \leq  K\, e^{-c(t-t_0)}\,\lvert x_0-y_0\rvert,\quad \forall t \geq t_0,
\end{equation}
that is, system \eqref{eq:dynamical_sys} is \emph{incrementally exponentially stable} ($IES$) in $\mathcal{C}$.
\end{thm}
In this paper we analyze contraction properties of dynamical systems based on norms and matrix measures \citep{lohmiller1998contraction,russo2010global}.
Other more general definitions exist in the literature, for example results based on Riemannian metrics \citep{lohmiller1998contraction} and Finsler-Lyapunov functions \citep{forni2014differential}.
The relationships between these three definitions and the definition of convergence \citep{pavlov2004convergent} were investigated in \citep{forni2014differential}.
\subsection{Switched systems}
\begin{figure}[!t]
\begin{center}
\includegraphics[width=2in]{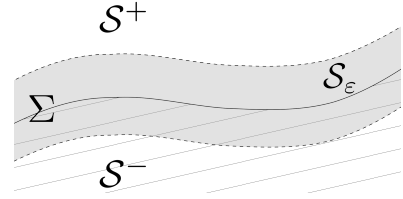} 
\caption{Regions of state space: switching manifold $\Sigma$, 
$\mathcal{S}^+$, $\mathcal{S}^-$ (hatched zone) and $\mathcal{S}_{\varepsilon}$ (grey zone).} 
\label{fig:regions}
\end{center}
\end{figure}

Switched (or bimodal) Filippov systems are dynamical systems $\dot{x}=f(x)$ where $f(x)$ is a piecewise continuous vector field having a codimension-one submanifold $\Sigma$ as its discontinuity set \citep{filippov1988differential,utkin2013sliding}.
The submanifold $\Sigma$ is called the \emph{switching manifold} and is defined as the zero set of a smooth function $h:\,U\subseteq \mathbb{R}^n\rightarrow\mathbb{R}$, that is
$
\Sigma:=\{x\in U : h(x)=0\},
$
where $0\in\mathbb{R}$ is a regular value of $h$, i.e. $\nabla h(x)\neq 0,\, \forall x\in\Sigma$.
$\Sigma$ divides $U$ in two disjoint regions, $\mathcal{S}^+:=\{x\in U : h(x)>0\}$ and ${\mathcal{S}^-:=\{x\in U : h(x)<0\}}$ (see Fig.~\ref{fig:regions}). \\
Hence, a bimodal Filippov system can be defined as
\begin{equation}
\label{eq:filippov_bimodal}
\dot{x}=
\begin{cases}
f^+(x), & \text{if } x\in\mathcal{S}^+ \\
f^-(x), & \text{if } x\in\mathcal{S}^-
\end{cases} ,
\end{equation}
where $f^+,f^-\in{C}^1(U,\mathbb{R}^n)$.
%
%
We assume that solutions of system \eqref{eq:filippov_bimodal} are defined in the sense of Filippov (and therefore admitting \emph{sliding} motions on $\Sigma$) and they have the property of \emph{right-uniqueness} in $U$ \cite[pag. 106]{filippov1988differential}.
Condition \eqref{eq:contraction_cond} was previously presented as a sufficient condition for a dynamical system to be incrementally exponentially stable, but it cannot be directly applied to system \eqref{eq:filippov_bimodal} because its vector field is not continuously differentiable.
In recent work reported in \citep{fiore2015contraction}, sufficient conditions were derived for convergence of any two trajectories of a Filippov system towards each other.
Instead of directly analyzing the Filippov vector field on $\Sigma$, the analysis is conducted on its regularization, say $f_\varepsilon(x)$,  defined as
\begin{equation*}
\label{eq:regularized_sys}
f_\varepsilon(x)=\frac{1+\varphi_\varepsilon \left( h(x) \right)}{2}\, f^+(x) +
\frac{1-\varphi_\varepsilon \left( h(x) \right)}{2}\, f^-(x),
\end{equation*}
where $\varphi_\varepsilon\in C^1(\mathbb{R},\mathbb{R})$ is the so-called transition function.
In this new system the switching manifold $\Sigma$ is replaced by a boundary layer $\mathcal{S}_\varepsilon$ (Fig.~\ref{fig:regions}) of width $2\varepsilon$, defined as
$
\mathcal{S}_{\varepsilon}:=\{x\in U : -\varepsilon <h(x)<\varepsilon \},
$
and, more importantly, $f_\varepsilon$ is continuously differentiable in $U$, so that condition \eqref{eq:contraction_cond} can be applied to it.  
Finally, contraction properties of Filippov systems \eqref{eq:filippov_bimodal} are recovered taking the limit for $\varepsilon\rightarrow 0$ and considering the following Lemma.%
\begin{lem}
\label{thm:regularization}
Denoting by $x_\varepsilon(t)$ a solution to the regularized system and by $x(t)$ a solution to the switched system with the same initial conditions $x_0$, then $\lvert x_\varepsilon(t)-x(t)\rvert=O(\varepsilon)$, uniformly for all $t\geq t_0$ and for all $x_0\in U$.
\end{lem}
For further details see \citep{sotomayor1996regularization, utkin2013sliding, fiore2015contraction}.
The resulting sufficient conditions for a bimodal Filippov system to be incrementally exponentially stable in a certain set are stated in the following theorem (see \citep{fiore2015contraction} for a complete proof and further details). 
\begin{thm}
\label{thm:contracting_pws}
The bimodal switched system \eqref{eq:filippov_bimodal} is incrementally exponentially stable in a $K$-reachable set $\mathcal{C}\subseteq U$ with convergence rate $c:=\min\,\{c_1,c_2\}$ if there exists some norm in $\mathcal{C}$, with associated matrix measure $\mu$, such that, for some positive constants $c_1,c_2$,
\begin{equation*}
\begin{aligned}
\mu\left( \frac{\partial f^+}{\partial x}(x)\right) & \leq -c_1, && \forall x \in \bar{\mathcal{S}}^+ ,\\
\mu\left( \frac{\partial f^-}{\partial x}(x)\right) & \leq -c_2,  && \forall x \in \bar{\mathcal{S}}^- ,\\ 
\mu\left( \Big[ f^+(x)-f^-(x)\Big] \nabla h(x) \right) & = 0,  && \forall x \in \Sigma.
\end{aligned}
\end{equation*}
\end{thm}
In the above relations $\bar{\mathcal{S}}^+$ and $\bar{\mathcal{S}}^-$ represent the closures of the sets $\mathcal{S}^+$ and $\mathcal{S}^-$, respectively.
%
%
\section{State observer design}
\label{sec:observer_design}
\subsection{Problem formulation}
Consider the bimodal switched system 
\begin{eqnarray}
\label{eq:bimodal_system_x}
&\dot{x}=
\begin{cases}
f^+(x)+u(t), & h(x)>0\\
f^-(x)+u(t), & h(x)<0
\end{cases}, \\
\label{eq:bimodal_system_y}
&y=g(x) ,
\end{eqnarray}
where $x\in\mathbb{R}^n$, $y\in\mathbb{R}^p$, $u\in\mathbb{R}^n$ are the state, output and the input of the system, respectively, and $f^+$, $f^-$, $g$ are continuously differentiable vector fields. \\
As an observer for the system \eqref{eq:bimodal_system_x}-\eqref{eq:bimodal_system_y}, we propose a bimodal Luenberger-like switched observer of the form
\begin{eqnarray}
\label{eq:observer_x}
&\dot{\hat{x}}=
\begin{cases}
f^+(\hat{x})+L^+(y-\hat{y})+u(t), & h(\hat{x})>0\\
f^-(\hat{x})+L^-(y-\hat{y})+u(t), & h(\hat{x})<0
\end{cases}, \\
\label{eq:observer_y}
&\hat{y}=g(\hat{x}),
\end{eqnarray}
where $\hat{x}(t)\in\mathbb{R}^n$ is the estimated state and $L^+,L^-\in\mathbb{R}^{n\times p}$ are  observer gain matrices to be selected appropriately. \\
We are interested in deriving conditions on the observer gain matrices $L^+$ and $L^-$ that guarantee exponential convergence to 0 of the estimation error $e(t):=x(t)-\hat{x}(t)$ for all $x(t): \mathbb{R}^+\rightarrow \mathbb{R}^n$ satisfying \eqref{eq:bimodal_system_x}-\eqref{eq:bimodal_system_y} for any given continuous function $u(t):\mathbb{R}^+\rightarrow \mathbb{R}^n$.
Note that in what follows we will not require system \eqref{eq:bimodal_system_x}-\eqref{eq:bimodal_system_y} to be contracting, i.e. Theorem \ref{thm:contracting_pws} must not necessarily hold for this system.
Instead, contraction theory will be exploited to analyze convergence of the system describing the dynamics of the estimation error.

\subsection{Main results}

\begin{thm}
\label{thm:observer}
The state estimation error $e(t)$ converges exponentially to zero, that is, for some $c>0$,
\begin{equation}
\label{eq:error_exp}
\left| e(t) \right|\leq K\,e^{-c(t-t_0)}\left| x(t_0) \right|,\ \quad \forall t\geq t_0,
\end{equation}
if there exists some matrix measure $\mu$, such that, for some positive constants $c_1$, $c_2$,
\begin{equation}
\label{eq:thm:condition1}
\mu\left( \frac{\partial f^+}{\partial x}(\hat{x}) - L^+ \frac{\partial g}{\partial x}(\hat{x}) \right) \leq -c_1, \forall \hat{x} : h(\hat{x})>0,
\end{equation}
\begin{equation}
\label{eq:thm:condition2}
\mu\left( \frac{\partial f^-}{\partial x}(\hat{x})- L^- \frac{\partial g}{\partial x}(\hat{x})\right) \leq -c_2,   \forall \hat{x} : h(\hat{x})<0,
\end{equation}
\begin{equation}
\label{eq:thm:condition3}
\mu\left( \Big[ \Delta f(\hat{x})+\Delta L (y-\hat{y})\Big] \nabla h(\hat{x}) \right) = 0,  \forall \hat{x} : h(\hat{x})=0,
\end{equation}
where $\Delta f(\hat{x})=f^+(\hat{x})-f^-(\hat{x})$ and $\Delta L=L^+ - L^-$.
Moreover, the convergence rate $c$ can be estimated as $\min\{c_1,c_2\}$.
\end{thm}

\begin{pf}
Conditions \eqref{eq:thm:condition1}-\eqref{eq:thm:condition3} come from the application of Theorem \ref{thm:contracting_pws} to the dynamics of the state observer \eqref{eq:observer_x}-\eqref{eq:observer_y} by rewriting them as
\begin{equation*}
\dot{\hat{x}}=
\begin{cases}
\bar{f}^+(\hat{x})+ \eta^+(t), & h(\hat{x})>0\\
\bar{f}^-(\hat{x})+ \eta^-(t), & h(\hat{x})<0
\end{cases},
\end{equation*}
where $\bar{f}^\pm(\hat{x})=f^\pm(\hat{x})-L^\pm g(\hat{x})$ depends only on $\hat{x}$, and $\eta^\pm(t)=L^\pm g(x(t))+u(t)$ is a function of $t$.\\
Hence, if such conditions are satisfied, then the state observer is contracting; this in turn implies that, for two generic solutions $\hat{x}_1(t)$ and $\hat{x}_2(t)$, \eqref{eq:ies} holds, i.e.
\begin{equation*}
|\hat{x}_1(t)-\hat{x}_2(t)|\leq K\, e^{-c(t-t_0)}|\hat{x}_1(t_0)-\hat{x}_2(t_0)|, \;\forall t\geq t_0.
\end{equation*}
Now, notice that a solution $x(t)$ of system \eqref{eq:bimodal_system_x} is a particular solution of the observer \eqref{eq:observer_x} --- because \eqref{eq:bimodal_system_x} and \eqref{eq:observer_x} have the same structure, except for the correction term $g(x) - g(\hat{x})$, which is null when considering $x(t)$ as a solution of the observer.
Then, we can replace $\hat{x}_2(t)$ with $x(t)$, rename $\hat{x}_1(t)$ as the general solution $\hat{x}(t)$, and write
\begin{equation*}
\begin{split}
|e(t)|= |x(t)-\hat{x}(t)|  \leq K\, e^{-c(t-t_0)}|x(t_0)|, 
\end{split}
\end{equation*}
for all $t\geq t_0$, where $\hat{x}(t_0)=0$ as usual in observer design.
Hence, the exponential convergence to zero of the estimation error is proved.
\end{pf}

\begin{remark}
Alternatively, the theorem can be proved considering the regularized dynamics of both system \eqref{eq:bimodal_system_x} and observer \eqref{eq:observer_x}.
Denoting by $x_\varepsilon(t)$ a solution to the regularized switched system \eqref{eq:bimodal_system_x}, and by $\hat{x}_\varepsilon(t)$ a solution to the regularized observer \eqref{eq:observer_x}, we have
\begin{equation*}
\begin{split}
|e(t)|=&|x(t)-\hat{x}(t)|\\
\leq & |x(t)-x_\varepsilon(t)|+|x_\varepsilon(t)-\hat{x}_\varepsilon(t)|+|\hat{x}_\varepsilon(t)-\hat{x}(t)|.
\end{split}
\end{equation*}
The first and the third terms are the error between a solution to the discontinuous system and a solution to its regularized counterpart; hence, from Lemma \ref{thm:regularization} we know that $|x(t)-x_\varepsilon(t)|=O(\varepsilon)$, and $|\hat{x}(t)-\hat{x}_\varepsilon(t)|=O(\varepsilon)$.\\
Furthermore, similarly to what done in \citep{fiore2015contraction}, it can be shown that conditions \eqref{eq:thm:condition1}-\eqref{eq:thm:condition3} imply incremental stability of the trajectories of the regularized observer, thus
\begin{equation*}
|\hat{x}_{\varepsilon,1}(t)-\hat{x}_{\varepsilon,2}(t)|\leq K\, e^{-c(t-t_0)}|\hat{x}_{\varepsilon,1}(t_0)-\hat{x}_{\varepsilon,2}(t_0)|, \forall t\geq t_0.
\end{equation*}
The theorem is finally proved by taking the limit for $\varepsilon\to 0^+$ and taking the same last step as that in the proof of Theorem \ref{thm:observer}.
\end{remark}
\begin{remark}
If one of the two modes, $f^+$ or $f^-$, of the observed system \eqref{eq:bimodal_system_x} is already contracting, the corresponding observer gain matrix, $L^+$ or $L^-$, in \eqref{eq:observer_x} can be set to zero to simplify the design problem.
The drawback is that the convergence rate of the estimation error will depend on that of the contracting mode that cannot be altered if this choice is made.
\end{remark}
\begin{remark}\label{rem:uncertainty}
In the presence of bounded disturbances or uncertainties on the models, contraction properties of the vector fields guarantee boundedness of the estimation error (a more detailed analysis is not the aim of the current paper; the interested reader can refer to \citep{lohmiller1998contraction}).
\end{remark}
%

\section{Examples}
\label{sec:examples}
\begin{figure}[!t]
\begin{center}
\begin{subfigure}[t]{0.03\linewidth}
a)
\end{subfigure}
\begin{subfigure}[t]{0.9\linewidth}
\includegraphics[width=\linewidth,valign=t]{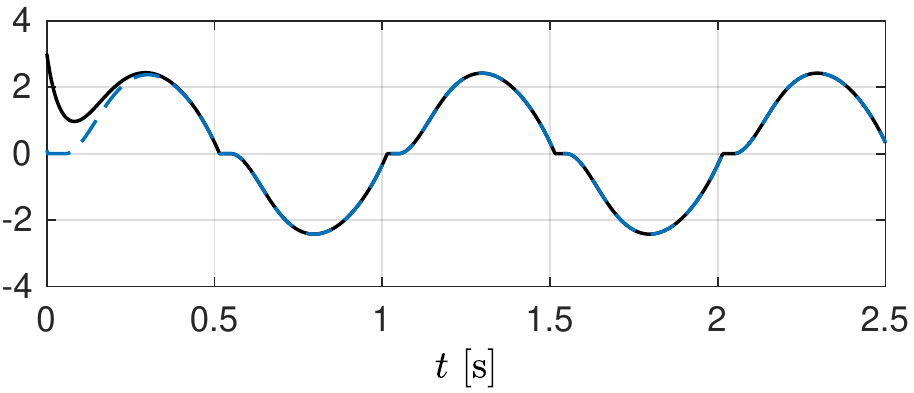}
\end{subfigure}\\
\begin{subfigure}[t]{0.03\linewidth}
b)
\end{subfigure}
\begin{subfigure}[t]{0.9\linewidth}
\includegraphics[width=\linewidth,valign=t]{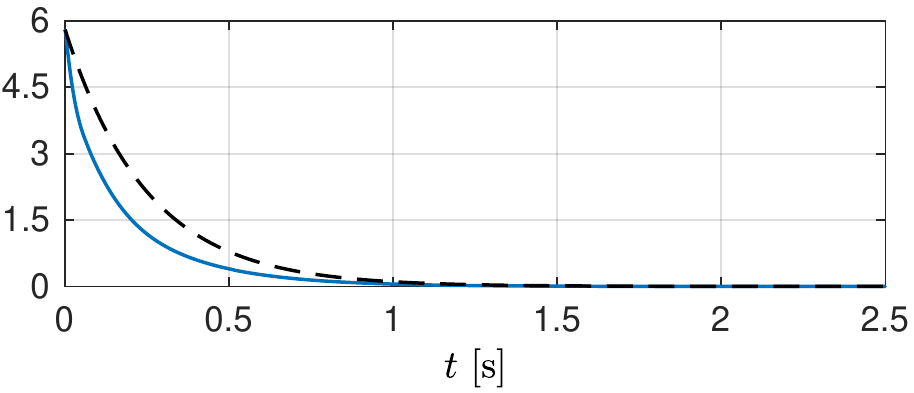}
\end{subfigure}\\
\caption{ {Panel a: Time evolution of the states $x_1(t)$ (solid line) and $\hat{x}_1(t)$ (dashed line) of Example 1, with initial conditions $x_0=[3\;\; 3]^{\mathrm{T}}$, $\hat{x}_0=[0\;\; 0]^{\mathrm{T}}$. Panel b: Norm of the estimation error $|e(t)|_1$. The dashed line represents the analytical estimate \eqref{eq:error_exp} with $c=4$ and $K=1$. Parameters: $L^+ =[-2\;\;0]^{\mathrm{T}}$ and $L^- =[2\;\;0]^{\mathrm{T}}$.}}
\label{fig:ex_pws}
\end{center}
\end{figure}

\begin{figure}[]
\begin{center}
\begin{subfigure}[t]{0.03\linewidth}
a)
\end{subfigure}
\begin{subfigure}[t]{0.9\linewidth}
\includegraphics[width=\linewidth,valign=t]{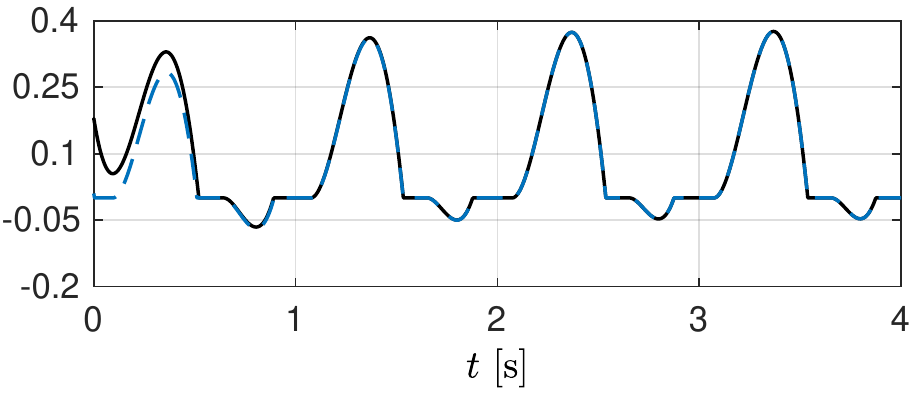}
\end{subfigure}\\
\begin{subfigure}[t]{0.03\linewidth}
b)
\end{subfigure}
\begin{subfigure}[t]{0.9\linewidth}
\includegraphics[width=\linewidth,valign=t]{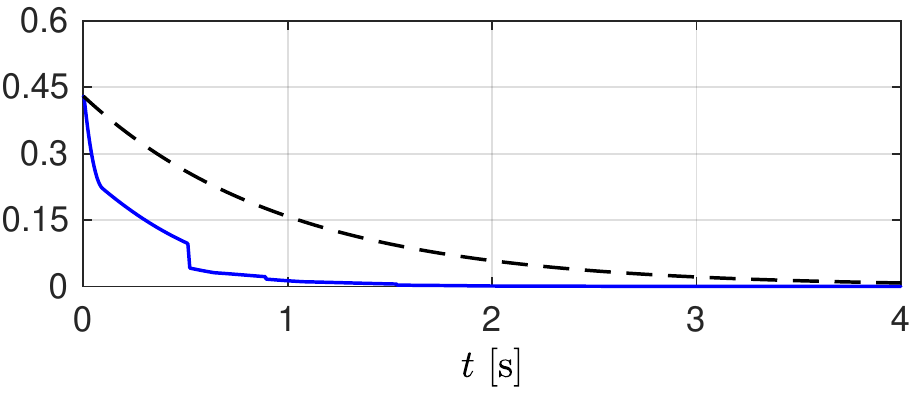}
\end{subfigure}\\
\begin{subfigure}[t]{0.03\linewidth}
c)
\end{subfigure}
\begin{subfigure}[t]{0.9\linewidth}
\includegraphics[width=\linewidth,valign=t]{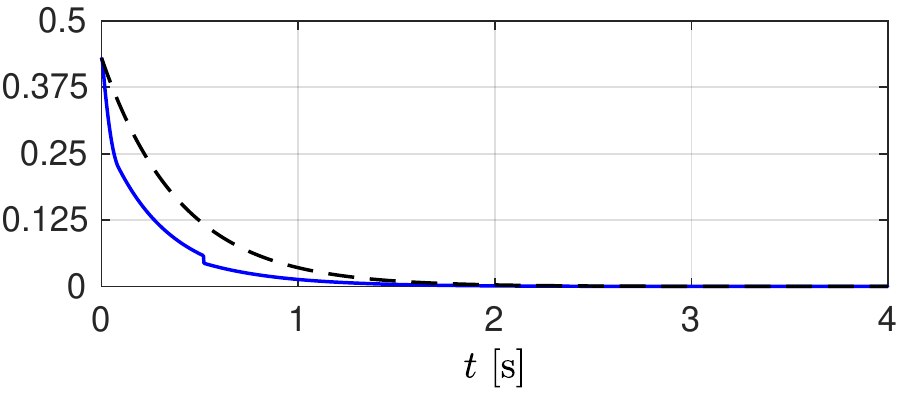}
\end{subfigure}\\
\caption{ {Panel a: Time evolution of the states $x_2(t)$ (solid line) and $\hat{x}_2(t)$ (dashed line) of Example 2, with initial conditions $x_0=[0.3\;\; 0.3]^{\mathrm{T}}$, $\hat{x}_0=[0\;\; 0]^{\mathrm{T}}$. Panel b: Norm of the estimation error $|e(t)|_1$. The dashed line represents the analytical estimate \eqref{eq:error_exp} with $c=1$ and $K=1$. Parameters: $L^+=L^-=[1\;\;1]^{\mathrm{T}}$. Panel c: Norm of the estimation error using observer gain $L=[1.5\;\;2]^{\mathrm{T}}$.}}
\label{fig:ex_pwa}
\end{center}
\end{figure}

\begin{figure}[!t]
\begin{center}
\begin{subfigure}[t]{0.03\linewidth}
a)
\end{subfigure}
\begin{subfigure}[t]{0.9\linewidth}
\includegraphics[width=\linewidth,valign=t]{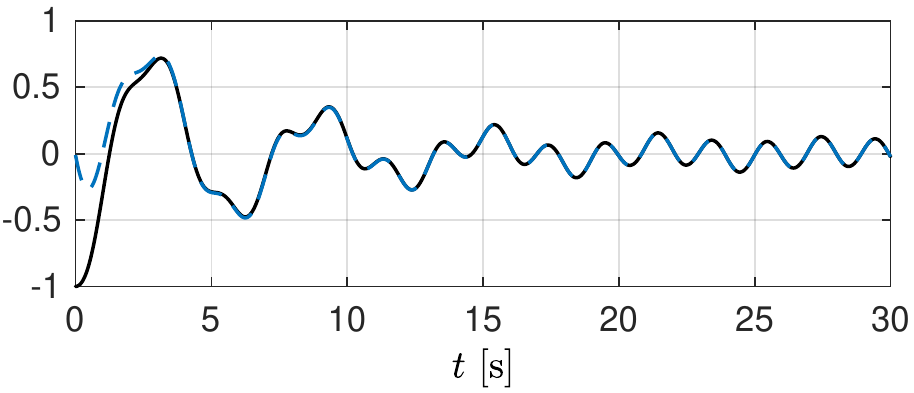}
\end{subfigure}\\
\begin{subfigure}[t]{0.03\linewidth}
b)
\end{subfigure}
\begin{subfigure}[t]{0.9\linewidth}
\includegraphics[width=\linewidth,valign=t]{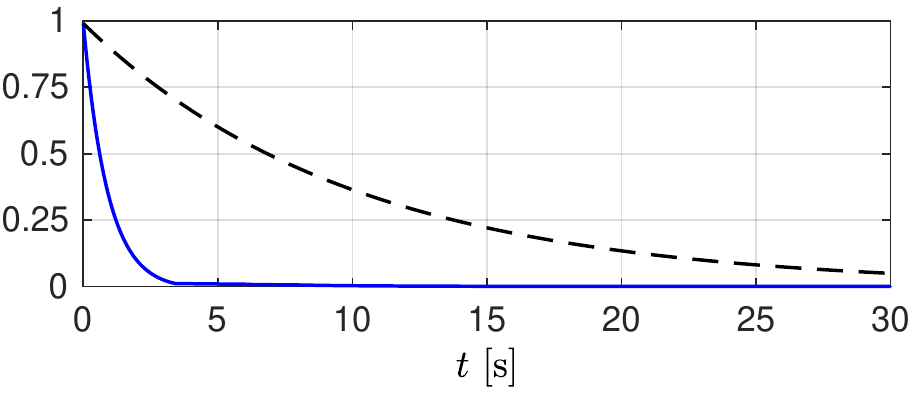}
\end{subfigure}\\
\caption{ {Panel a: time evolution of the states $x_1(t)$ (solid line) and $\hat{x}_1(t)$ (dashed line) of Example 3, with initial conditions $x_0=[-1\;\; 0]^{\mathrm{T}}$, $\hat{x}_0=[0\;\; 0]^{\mathrm{T}}$. Panel b: Norm of the estimation error $|e(t)|_\infty$. The dashed line represents the analytical estimate \eqref{eq:error_exp} with $c=0.1$ and $K=1$. Parameters: $\omega_{\mathrm{n}} = 1 \ \mathrm{rad/s}$, $Q = 10$, $m = 1 \ \mathrm{kg}$, $F_{\mathrm{d}} = 1 \ \mathrm{N}$, $\omega_{\mathrm{d}}= \pi \ \mathrm{rad/s}$, $F_{\mathrm{f}} = 0.1 \ \mathrm{N}$,  $\ell_1 = 1.1$, $\ell_2 = -1$.}}
\label{fig:ex_df}
\end{center}
\end{figure}

Here we present some examples to illustrate the use of Theorem \ref{thm:observer} for the design of observers for switched systems.
All simulations presented in this section have been computed using the numerical solver in \citep{piiroinen2008event}.
\paragraph*{Example 1} 
Consider a nonlinear bimodal switched system as in \eqref{eq:bimodal_system_x}-\eqref{eq:bimodal_system_y} with 
\begin{equation*}
f^+ (x)\! =\! \begin{bmatrix}
-9 x_1 - 3x_1^2 - 18\\
-4 x_2
\end{bmatrix}\!,
\,
f^- (x)\! =\! \begin{bmatrix}
-9 x_1 + 3x_1^2 + 18\\
-4 x_2
\end{bmatrix}\!,
\end{equation*}
and $h (x) = x_1$, $y=g (x) = x_1^2$.\\
According to Theorem \ref{thm:observer}, a state observer as in \eqref{eq:observer_x}-\eqref{eq:observer_y} with $L^+ = [ \ell_1^+  \;\;  \ell_2^+]^\mathrm{T}$ and $L^- = [\ell_1^- \;\;  \ell_2^-]^\mathrm{T}$ for this system has the property that its estimation error converges exponentially to zero if there exist choices of the gain matrices $L^+$ and $L^-$ so that all three conditions \eqref{eq:thm:condition1}-\eqref{eq:thm:condition3} are satisfied. \\
%
To find  $L^+$ and $L^-$, it is first necessary to select a specific matrix measure; here we use the measure $\mu_1$, associated to the so-called $\ell^1$-norm (see Appendix \ref{sec:appendix}). Therefore, conditions \eqref{eq:thm:condition1} and \eqref{eq:thm:condition2} translate respectively to
\begin{equation*}
\mu_1 \left(
 \begin{bmatrix}
-9 - 6 \hat{x}_1 - 2 \ell_1^+ \hat{x}_1  &  0\\
- 2 \ell_2^+ \hat{x}_1  &  -4
\end{bmatrix}
 \right) < 0, \quad \text{with } \hat{x}_1 >0 ,
\end{equation*}
\begin{equation*}
\mu_1 \left(
 \begin{bmatrix}
-9 + 6 \hat{x}_1 - 2 \ell_1^- \hat{x}_1  &  0\\
- 2 \ell_2^- \hat{x}_1  &  -4
\end{bmatrix}
 \right) < 0, \quad \text{with } \hat{x}_1 <0.
\end{equation*}
Selecting for simplicity $\ell_2^+ = \ell_2^- = 0$, the above inequalities are satisfied if
\begin{equation*}
\max \{-9 - 6 \hat{x}_1 - 2 \ell_1^+ \hat{x}_1; \; -4\} < 0, \quad \text{with } \hat{x}_1 > 0,
\end{equation*}
\begin{equation*}
\max \{-9 + 6 \hat{x}_1 - 2 \ell_1^- \hat{x}_1; \; -4 \} < 0, \quad \text{with } \hat{x}_1 < 0.
\end{equation*}
This is true if $\ell_1^+ > -3$ and $\ell_1^- < 3$.\\
Next, from the the third condition \eqref{eq:thm:condition3}, we have 
\begin{equation*}
\begin{split}
\mu_1 \left(
 \begin{bmatrix}
- 6 \hat{x}_1^2 - 36 + (\ell_1^+ - \ell_1^-)(x_1^2 - \hat{x}_1^2)\\
0
\end{bmatrix}
\begin{bmatrix}
1  &  0 \\
\end{bmatrix}
 \right) = 0,
\end{split}
\end{equation*}
with $\hat{x}_1 = 0$, which is verified if
$
\max \{ - 36 + (\ell_1^+ - \ell_1^-)x_1^2; \; 0\} = 0,
$
 i.e. if
$
- 36 + (\ell_1^+ - \ell_1^-)x_1^2 < 0,
$
which holds for all $x_1$ if $\ell_1^+ < \ell_1^-$.
Therefore, to satisfy all three conditions of Theorem \ref{thm:observer}, it is possible for example to select $L^+ =[-2\;\;0]^{\mathrm{T}}$ and $L^- =[2\;\;0]^{\mathrm{T}}$. The resulting state observer is contracting and its estimation error satisfies \eqref{eq:error_exp} with convergence rate $c=4$.
%
In Fig.~\ref{fig:ex_pws}(a) we show numerical simulations of the evolution of the states $x_1$ and $\hat{x}_1$ when an input $u(t)=[1\;\;1]^{\mathrm{T}}\, \sin(2\pi t)$ of period $T=1$ is applied to the system. In Fig.~\ref{fig:ex_pws}(b) the evolution of the $\ell^1$-norm of the state estimation error $e(t)$ is reported, confirming the analytical estimate \eqref{eq:error_exp}.
%
\paragraph*{Example 2}
Consider a piecewise affine (PWA) system of the form
\begin{eqnarray}
\label{eq:pwa_sys}
&\dot{x}=
\begin{cases}
A_1x+b_1+Bu, & \text{if } h^{\mathrm{T}}x>0\\
A_2x+b_2+Bu, & \text{if } h^{\mathrm{T}}x<0
\end{cases},\\
&y=c^{\mathrm{T}}x,
\end{eqnarray}
where 
\begin{equation*}
A_1=
\begin{bmatrix}
-1 & 0\\
2 & -2
\end{bmatrix},
\,
b_1=
\begin{bmatrix}
-1\\
-3
\end{bmatrix},
\,
A_2=
\begin{bmatrix}
-1 & 0\\
2 & -3
\end{bmatrix},
\,
b_2=
\begin{bmatrix}
2\\
4
\end{bmatrix},
\end{equation*}
and $B=[0\;\;1]^{\mathrm{T}}$, $h=[0\;\;1]^{\mathrm{T}}$, $c=[1\;\;1]^{\mathrm{T}}$.\\
A state observer as in \eqref{eq:observer_x}-\eqref{eq:observer_y} for this system has the structure
\begin{eqnarray}
\label{eq:pwa_observer}
&\dot{\hat{x}}=
\begin{cases}
A_1\hat{x}+b_1+L^+(y-\hat{y})+Bu, & \text{if } h^{\mathrm{T}}\hat{x}>0\\
A_2\hat{x}+b_2+L^-(y-\hat{y})+Bu, & \text{if } h^{\mathrm{T}}\hat{x}<0
\end{cases},\\
&\hat{y}=c^{\mathrm{T}}\hat{x},
\end{eqnarray}
where, for the sake of simplicity, we choose $L^+ = L^- = L$.
Again we decide to proceed using the matrix measure induced by the $\ell^1$-norm. In this case, conditions \eqref{eq:thm:condition1} and \eqref{eq:thm:condition2} yield respectively $\mu_1 \left( A_1-Lc^{\mathrm{T}} \right)= \max\{ -1-\ell_1+|2-\ell_2|;\, -2-\ell_2+|\ell_1| \}$, and $\mu_1 \left( A_2-Lc^{\mathrm{T}} \right)=\max\{ -1-\ell_1+|2-\ell_2|;\, -3-\ell_2+|\ell_1| \}$.
%
It is easy to verify that choosing $\ell_1=\ell_2=1$ both measures are equal to $-1$. 
Condition \eqref{eq:thm:condition3} is  verified independently of $L$.\\
Hence, the designed observer \eqref{eq:pwa_observer} is contracting and the estimation error converges exponentially to zero with rate $c=1$. In Fig.~\ref{fig:ex_pwa}(a) we show numerical simulations of the evolution of the states $x_2$ and $\hat{x}_2$ when an input $u(t)=4\,\sin (2\pi t)$ of period $T=1$ is applied to the system. In Fig.~\ref{fig:ex_pwa}(b)  the evolution is reported of the $\ell^1$-norm of the state estimation error $e(t)$.\\
Note that faster convergence can be obtained by choosing higher values of $\ell_1$ and $\ell_2$ fulfilling conditions (25)-(26). For example choosing $L=[1.5\;\;2]^T$ we obtain a convergence rate $c=2.5$, as shown in Fig.~\ref{fig:ex_pwa}(c).
%
%
\paragraph*{Example 3}
Consider now a harmonic oscillator affected by Coulomb friction, described by the equations
\begin{eqnarray}
\label{eq:oscillator_x}
&\begin{cases}
\dot{x}_1 =  x_2,  \\
\dot{x}_2 =  - \omega_{\mathrm{n}} x_1 - \dfrac{\omega_{\mathrm{n}}}{Q} x_2 - \dfrac{F_{\mathrm{f}}}{m} \mathrm{sgn} (x_2) + \dfrac{F_{\mathrm{d}}}{m} \sin (\omega_{\mathrm{d}} t) , \\
\end{cases} \\
\label{eq:oscillator_y}
&y =  x_1,
\end{eqnarray}
where $x_1 \in \mathbb{R}$ is the position of the oscillator, $x_2 \in \mathbb{R}$ is its velocity, $\omega_{\mathrm{n}}$ is its natural frequency, $Q$ is said \textit{Q factor} and is inversely proportional to the damping, $m$ is the mass of the oscillator, $F_{\mathrm{d}}$ is the amplitude of the driving force, $\omega_{\mathrm{d}}$ is the driving frequency and $F_{\mathrm{f}}$ is the  amplitude of the dry friction force which is modeled through the sign function as in \citep{csernak2006periodic}.
The proposed observer for system \eqref{eq:oscillator_x}-\eqref{eq:oscillator_y} has the form
\begin{equation*}
\begin{cases}
\dot{\hat{x}}_1 =  \hat{x}_2 + \ell_1 ( x_1 - \hat{x}_1) \\
\begin{split}
\dot{\hat{x}}_2 =  - \omega_{\mathrm{n}} \hat{x}_1 - \dfrac{\omega_{\mathrm{n}}}{Q} \hat{x}_2  -\dfrac{F_{\mathrm{f}}}{m} \mathrm{sgn} (\hat{x}_2) \\
 + \ell_2 ( x_1 - \hat{x}_1) + \dfrac{F_{\mathrm{d}}}{m} \sin (\omega_{\mathrm{d}} t)
\end{split}
\end{cases},
\;\; \hat{y} = \hat{x}_1.
\end{equation*}
Note that system \eqref{eq:oscillator_x} may be viewed as a PWA system \eqref{eq:pwa_sys} where
\begin{equation*}
A_1=A_2=
\begin{bmatrix}
0 & 1\\
-\omega_{\mathrm{d}} & -\omega_{\mathrm{d}}/Q
\end{bmatrix},
\end{equation*}
$B=[0\;\; 1/m]^{\mathrm{T}}$, $b_1=[0\;\; -F_{\mathrm{f}}/m]^{\mathrm{T}}$, $b_2=[0 \;\; F_{\mathrm{f}}/m]^{\mathrm{T}}$, $h=[0\;\;1]^{\mathrm{T}}$, and excited by an input $u(t)=F_\mathrm{d}\sin (\omega_{\mathrm{d}} t)$.\\
%
Using the measure $\mu_\infty$ induced by the uniform norm (see Appendix \ref{sec:appendix}), conditions \eqref{eq:thm:condition1} and \eqref{eq:thm:condition2} of Theorem \ref{thm:observer}, combined, yield
\begin{equation*}
\mu_\infty \left(
  \begin{bmatrix}
- \ell_1  &  1\\
- \omega_{\mathrm{n}} - \ell_2  &  -\omega_{\mathrm{n}}/Q
\end{bmatrix}
 \right) < 0, \quad \text{with } \hat{x}_2 \neq 0 ,
\end{equation*}
which in turn is equivalent to
\begin{equation*}
\max \left\{-\ell_1 + 1; \; -\omega_{\mathrm{n}}/Q+\left| -\omega_{\mathrm{n}}-\ell_2 \right|\right\} < 0, \quad \text{with } \hat{x}_2 \neq 0.
\end{equation*}
Therefore $\ell_1$ and $\ell_2$ must be chosen so that $\ell_1 >  1$, and $-\omega_{\mathrm{n}} \left( 1 +{1}/{Q} \right) < \ell_2 <  -\omega_{\mathrm{n}} \left( 1 -{1}/{Q} \right)$.\\
Furthermore, condition \eqref{eq:thm:condition3} is verified if
$
\max\! \left\{ 0; -F_{\mathrm{f}}/m \right\} = 0$,
which always holds because $F_{\mathrm{f}}, m > 0$.\\
Numerical simulations reported in Fig.~\ref{fig:ex_df}(a)-(b) confirm the theoretical predictions, showing that the estimation error converges to zero.
In practice, the exact value of the parameter $F_f$ is not known.
This implies bounded convergence of the estimation error, as stated in Remark \ref{rem:uncertainty}.
%
%
%
\section{Conclusions}
\label{sec:conclusion}
We presented an approach based on contraction for the design of state observers for a large class of nonlinear switched systems including those exhibiting sliding motion, such as the friction oscillator. The design methodology is based on the analysis of incremental exponential stability based on the extension of contraction theory to switched bimodal Filippov systems derived in \citep{fiore2015contraction}. The conditions were formulated in terms of matrix measures of the Jacobians of the observer dynamics and of an additional condition on the vector fields on the discontinuity set. The theoretical results were illustrated through simple but representative examples demonstrating the effectiveness of the proposed methodology. 
Future work will be aimed at extending the approach to a wider class of switched systems, investigating constructive methods to design both metrics and observer gains, and reformulating the design procedure as a convex optimization problem to compute them numerically.
%
%
\bibliography{refs}            
%
\appendix
\section{Matrix measures}    
\label{sec:appendix}                                                                        %
%
The \emph{matrix measure} \citep{dahlquist1958stability,vidyasagar2002nonlinear} associated to a matrix $ A \in\mathbb{R}^{n \times n}$ is the function $\mu(\cdot):\mathbb{R}^{n \times n}\rightarrow \mathbb{R}$ defined as
$
\mu( A )=\lim_{h \rightarrow 0^+}\left({\lVert I+h A  \rVert-1}\right)/{h}.
$
%
See \citep{vidyasagar2002nonlinear,desoer1972measure} for a list of properties of this measure.
The most common used matrix measures are those associated to the $\ell^1$-norm, the Euclidean norm and the uniform norm, and they are defined as follows:
\begin{equation*}
\label{eq:mu_1}
\mu_{1}( A )=\max_{j}\bigg[a_{jj}+\sum_{i\ne j}|a_{ij}|\bigg],
\end{equation*}
%
$$
\mu_{2}( A )=\lambda_{max}\left(\frac{ A + A^{\mathrm{T}} }{2}\right),
$$ 
$$
\mu_{\infty}( A )=\max_{i}\bigg[a_{ii}+\sum_{j\ne i}|a_{ij}|\bigg].
$$
\end{document}